\documentclass[sigconf]{acmart}

\usepackage{booktabs}
\usepackage{multirow}
\usepackage{array}
\usepackage{amsmath}
\usepackage{xspace}
\usepackage{pifont}
\usepackage{pgfplots}
\pgfplotsset{compat=1.18}
\usepackage{tikz}

\copyrightyear{2026}
\acmYear{2026}
\setcopyright{rightsretained}
\acmConference[ASE '26]{41st IEEE/ACM International Conference on
  Automated Software Engineering}{October 12--16, 2026}{Munich, Germany}
\acmBooktitle{Proceedings of the 41st IEEE/ACM International Conference on
  Automated Software Engineering (ASE '26)}
\acmDOI{10.1145/nnnnnnn.nnnnnnn}
\acmISBN{978-x-xxxx-xxxx-x/YYYY/MM}

\begin{document}

\title{Log-Insight: Automating Microservice Incident Diagnosis via
  Neuro-Symbolic Log Analysis}

\author{Carlos Garcia-Hernandez}
\authornote{Equal contribution.}
\affiliation{%
  \institution{Huawei Ireland Research Centre}
  \city{Dublin}
  \country{Ireland}
}
\author{Aymane Abdali}
\authornotemark[1]
\affiliation{%
  \institution{Huawei Ireland Research Centre}
  \city{Dublin}
  \country{Ireland}
}
\author{Guangyu Wu}
\affiliation{%
  \institution{Huawei Ireland Research Centre}
  \city{Dublin}
  \country{Ireland}
}
\author{Mingxue Wang}
\affiliation{%
  \institution{Huawei Ireland Research Centre}
  \city{Dublin}
  \country{Ireland}
}
\author{Fei Shen}
\affiliation{%
  \institution{Huawei Dongguan R\&D Centre}
  \city{Dongguan}
  \country{China}
}
\author{Zhaoyu Pang}
\affiliation{%
  \institution{Huawei Dongguan R\&D Centre}
  \city{Dongguan}
  \country{China}
}
\author{Yanbin Zhang}
\affiliation{%
  \institution{Huawei Dongguan R\&D Centre}
  \city{Dongguan}
  \country{China}
}

\renewcommand{\shortauthors}{Garcia-Hernandez, Abdali, Wu, Wang, Shen, Pang, Zhang}

\begin{abstract}
Diagnosing production incidents in large-scale microservice systems is
labour-intensive and time-critical for Site Reliability Engineers (SREs).
A single 30-minute incident window in our deployment can generate over
two million log lines across many interdependent services---approximately
1.2 billion characters, or 26{,}000$\times$ the 46{,}000-character per-request
budget of our enterprise LLM API---making direct LLM-based Root Cause
Analysis (RCA) infeasible. Existing approaches each leave a gap:
template-based parsers compress volume but produce no semantic anomaly
reasoning; deep-learning anomaly detectors emit black-box binary signals;
and LLM-based pipelines suffer context overflow and domain hallucination
on raw telemetry.

We present \textsc{Log-Insight}, an automated incident-diagnosis system
deployed in production at Huawei. The key design principle is to automate
the SRE's manual triage workflow: symbolic stages replicate the structured
investigation a skilled SRE would perform---sampling, schema understanding,
pattern clustering, statistical anomaly ranking---and hand the LLM only a
compact, pre-ranked evidence dossier to synthesise into a ranked hypothesis
report. A six-stage pipeline (Two-Pass Sampling, Schema Inference \& Memory,
Drain3 Pattern Clustering, Two-Layer Entropy-Guided Compression, Contrastive
Skew Analysis, Generative Synthesis) reduces millions of raw events by
1{,}000--7{,}000$\times$ while preserving statistically significant failure
signals.

Evaluated on 11 historical production incidents (110 runs, SRE-validated
ground truth), \textsc{Log-Insight} achieves MRR~$=$~0.790, returning the
correct root cause within the top-3 hypotheses in over 90\% of runs in
under a minute of end-to-end latency. We report two systematic failure modes, two
active mitigations, and four open research directions. The Forensic
Evidence section---listing exact log templates and skew statistics---was
consistently identified by operators as a key adoption factor, shifting
the system's perceived role from opaque oracle to investigative assistant.
\end{abstract}



\maketitle

\section{Introduction}

The automation of software operations---monitoring, diagnosis, and
remediation---is increasingly central to automated software engineering.
As microservice architectures become the dominant deployment model for
large-scale software systems, incident management has emerged as one of
the most labour-intensive and error-prone tasks faced by operations
teams. SREs are expected to diagnose complex failures under strict time
pressure, yet the telemetry they must reason over---system logs, metrics,
traces---grows in volume faster than any team can manually inspect.

Log analysis is the cornerstone of incident diagnosis. Yuan et\,al.\
report that for a majority (84\%) of failures in distributed systems,
all triggering events are present in the logs~\cite{yuan2014simple}, and
log inspection is widely regarded as the first and most
time-consuming step in triage. The challenge is not the absence of data
but its overwhelming volume: in our production environment at Huawei, a
single 30-minute incident window generates over two million log lines
across dozens of interdependent microservices---approximately 1.2 billion
characters, or 26{,}000$\times$ the 46{,}000-character budget of our
deployed LLM API.

Three classes of automated approaches exist, each with a significant gap:
\begin{itemize}
  \item \textbf{Template-based parsers} (Drain~\cite{he2017drain},
    Spell~\cite{du2016spell}) efficiently reduce log volume through
    unsupervised clustering but produce no semantic anomaly
    interpretation: they identify which patterns exist, not why they
    are anomalous.
  \item \textbf{Deep-learning anomaly detectors}
    (DeepLog~\cite{du2017deeplog}, LogRobust~\cite{zhang2019robust})
    detect deviations from normal behaviour but emit black-box binary
    signals and cannot generate the actionable, natural-language RCA
    hypotheses that SREs can act upon during a live incident.
  \item \textbf{LLM-based pipelines}~\cite{chen2024rcacopilot,li2025coca,roy2024exploring}
    offer strong reasoning capabilities but suffer from context-window
    overflow, domain-specific hallucination, and the
    ``Lost in the Middle'' phenomenon~\cite{liu2024lost}, where models
    fail to attend to evidence injected deep within long prompts.
\end{itemize}

\paragraph{Our approach.}
\textsc{Log-Insight} bridges these gaps by automating the structured
investigation that a skilled SRE performs manually. Rather than feeding
raw logs to an LLM, a neuro-symbolic pipeline replicates the SRE's
cognitive process: sample representatively, understand the schema,
cluster recurring patterns, score anomalies statistically, and only
then ask the LLM to synthesise a ranked explanation from the
pre-digested evidence. Symbolic stages perform anomaly detection and
ranking; the LLM acts as a constrained synthesiser. Because the
LLM synthesises from statistically pre-ranked evidence rather than
raw telemetry, confabulation requires actively contradicting an
injected statistic---a substantially harder failure mode than
inferring from vague text.

\paragraph{Contributions.}
\begin{enumerate}
  \item A production-deployed neuro-symbolic incident-diagnosis system
    with a six-stage pipeline that achieves up to 7{,}000$\times$ log
    compression while preserving diagnostic signal (\S\ref{sec:system}).
  \item An industrial evaluation on 11 real production incidents (110
    runs, SRE-validated ground truth), achieving MRR~$=$~0.790 and
    top-3 accuracy in over 90\% of runs in under a minute of latency
    (\S\ref{sec:eval}).
  \item Analysis of two systematic failure modes (context omission
    and synthesis failure), two active mitigations, and qualitative
    evidence for the adoption-driving role of statistical
    transparency.
  \item Four open research directions generalisable beyond our
    deployment context (\S\ref{sec:limitations}).
\end{enumerate}

\section{Background and Related Work}

\subsection{Automated Log Analysis}
\textbf{Template-based parsing} (Drain~\cite{he2017drain},
Spell~\cite{du2016spell}) converts unstructured log messages into
structured event templates via unsupervised clustering. Drain uses
a fixed-depth parse tree for efficient online extraction and has
become the de-facto standard preprocessing step in large-scale AIOps
pipelines. Follow-on work has applied LLMs to improve parsing
accuracy~\cite{zhong2024logparser}, but all template-based methods
share a fundamental limitation: they produce no interpretation of
\emph{why} a cluster is anomalous, only that it exists.

\textbf{Deep-learning anomaly detection}
(DeepLog~\cite{du2017deeplog}, LogRobust~\cite{zhang2019robust})
treats log sequences as execution paths and flags deviations from
learned norms. DeepLog uses LSTMs to model normal system execution
and detect anomalous log keys; LogRobust improves robustness to log
format evolution via attention-based semantic encoding. These methods
are effective for binary anomaly detection but face two limitations
in production incident diagnosis: they cannot produce the
natural-language RCA reports that SREs require during live incidents,
and they degrade when log formats evolve, requiring frequent
retraining. More recent explanation-focused
models~\cite{zhang2025llmlade} improve interpretability but still
struggle with the raw-telemetry-volume problem.

\textbf{LLM-based log analysis and RCA.}
RCACopilot~\cite{chen2024rcacopilot} applies
in-context learning to structure cloud-incident diagnosis,
pre-processing incident data into structured summaries before LLM
reasoning. COCA~\cite{li2025coca} integrates code-level knowledge
into distributed-system RCA. Roy et\,al.~\cite{roy2024exploring}
explore multi-step LLM agents for RCA using tool-calling to retrieve
relevant evidence. All of these approaches apply LLMs to rich but
already-compressed contextual inputs produced by humans or
domain-specific preprocessing---they do not address the fundamental
problem of ingesting arbitrary raw telemetry at massive scale.

\subsection{Recent AIOps Systems}
LogPilot~\cite{jiang2025logpilot} introduces an intent-aware framework
for alert diagnosis that interprets PromQL alert definitions to scope
relevant logs, reconstructs spatiotemporal log chains, and clusters
chains to identify recurring patterns before LLM synthesis. Deployed
on Volcano Engine Cloud and evaluated on real-world alerts, it
reports a 50.3\% improvement in root-cause-summary usefulness and a
54.8\% improvement in localisation Exact Match over prior baselines.
LogSage~\cite{xu2025logsage} presents an end-to-end LLM framework for
CI/CD failure detection and remediation validated at ByteDance scale
(over 1.07M executions in a year-long deployment, with end-to-end
precision exceeding 80\%), using token-efficient log preprocessing
and RAG-based solution generation. TrioXpert~\cite{sun2025trioxpert}
automates the full incident-management workflow for microservice
systems including detection, localisation, and triage, leveraging
multimodal data.

\textsc{Log-Insight} is complementary in its input assumptions. LogPilot
and LogSage both assume structured, format-specific log inputs with
defined alert signals (PromQL expressions, CI/CD pipeline outputs).
\textsc{Log-Insight} targets a more general setting: arbitrary
structured log tables with heterogeneous schemas, no pre-defined
alert format, and extreme volume (up to 3.54\,M rows in a single
log space, or up to 398 columns in another). The entropy-based
context compression and schema-agnostic design make
\textsc{Log-Insight} applicable across diverse microservice domains
without format-specific engineering.

The Ericsson deployment study by Shahedi
et\,al.~\cite{shahedi2025excellence} provides important context for
our adoption findings. They identify what they call the
``Excellence Paradox'': technical sophistication can actively impede
adoption when it conflicts with usability and practitioner trust.
Our finding that statistical transparency---the Forensic Evidence
section---is a key adoption factor is a concrete instantiation of
this principle in the RCA domain.

\subsection{Entropy-Based Feature Selection and Context Management}
Stage~4 of \textsc{Log-Insight} applies Shannon entropy as a feature
selection criterion over log metadata columns, connecting to classical
mutual-information filtering~\cite{guyon2003introduction} and
information-theoretic feature ranking in data mining. While
entropy-based dimensionality reduction is well-studied in machine
learning, its application as a \emph{context-budget management
mechanism} for LLM-based analysis pipelines is, to our knowledge,
new. The key insight
is that fitting a diagnostic signal into a fixed-size context window
is structurally equivalent to selecting maximally informative features
under a budget constraint---a connection that makes existing feature-selection
methods directly applicable to LLM context engineering.

Liu et\,al.~\cite{liu2024lost} demonstrate that LLMs attend poorly to
relevant evidence in the middle of long contexts, motivating our
strict priority ordering in Forensic Case File assembly.
\textsc{Log-Insight}'s design---always placing the highest-signal
content (Critical Hints) at the beginning of the context---is a
direct architectural response to this empirical finding.

\section{System Architecture}
\label{sec:system}

\begin{figure}[h]
  \centering
  \includegraphics[width=\linewidth]{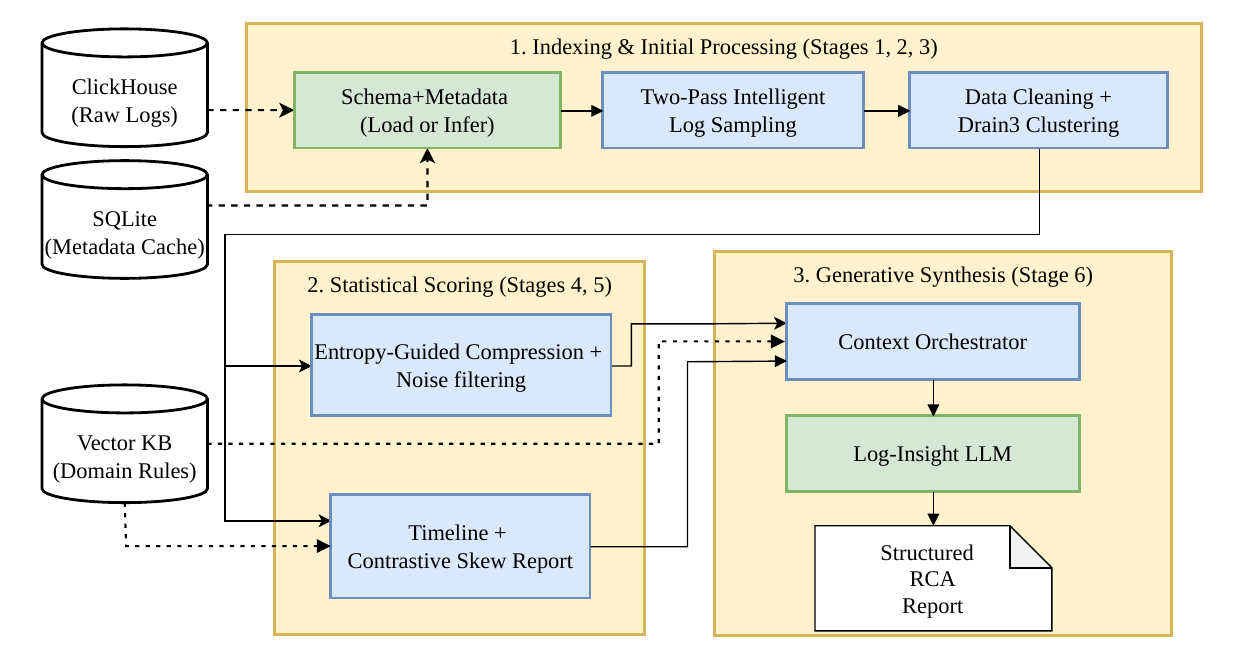}
  \caption{\textsc{Log-Insight} architecture. Symbolic stages handle
all high-volume data processing; the LLM acts only as a synthesiser
over the compressed context. The Knowledge Base feeds domain
rules into both the schema stage and the orchestrator.}
  \label{fig:arch}
\end{figure}

\textsc{Log-Insight} implements a six-stage pipeline (Figure~\ref{fig:arch})
that strictly separates high-volume symbolic preprocessing from neural
synthesis. Table~\ref{tab:trace} shows the system's exact internal
trace on a real production incident.

\begin{table}[t]
\caption{Pipeline trace on a representative 1{,}000-log production
incident.}
\label{tab:trace}
\small
\begin{tabular}{@{}p{1.7cm}p{6.0cm}@{}}
\toprule
\textbf{Stage} & \textbf{Transformation \& Engine Output} \\
\midrule
1.~Raw Input &
\texttt{\{"timestamp":"2026-01-14T03:22:11Z",
"flag":"F", "return\_code":"990000",
"hostAddr":"10.140.111.13"\}} \\
\midrule
2--3.~Schema \& KB &
\emph{SRE Rule loaded:} \texttt{flag='F'} $\Rightarrow$ failure;
\texttt{return\_code='990000'} $\Rightarrow$ ``Database Timeout''. \\
\midrule
4.~Entropy &
\texttt{timestamp} ($H{=}9.96$ bits) $\rightarrow$ abstracted (ID).
\texttt{return\_code} ($H{=}1.56$ bits) $\rightarrow$ ENUM retained.
\texttt{Pos0=NUM(0.1/0.1/0.2) | Pos1=CONST('tm-webcast')}. \\
\midrule
5.~Skew &
\emph{Hint 1:} \texttt{hostAddr='10.140.111.13'}: 15.3\% errors,
0\% successes.
\emph{Hint 2:} \texttt{clientVersion='12.11.30.303'}: 26.5\% errors,
0\% successes. \\
\midrule
6.~LLM Output &
``Systemic failure on host range 10.140.111.13--20; version
12.11.30.303 responsible for 26.5\% of errors.'' \\
\bottomrule
\end{tabular}
\end{table}

\subsection{Stage 1: Two-Pass Sampling}
Processing billions of raw log characters from ClickHouse is infeasible
within a 60-second SLA. \textsc{Log-Insight} implements a two-pass
sampling strategy that balances breadth and depth. The first pass
retrieves a random $10^4$-log seed
$\mathcal{L}_1 \sim \mathrm{Uniform}(\mathcal{L}, 10^4)$ to map the
schema and attribute distribution. The second pass performs targeted,
error-weighted oversampling. Samples are drawn across unique
\emph{structural fingerprints}---log lines with all variable tokens
masked to a fixed placeholder---ensuring rare error templates are not
crowded out by high-frequency nominal messages. If explicit severity
levels are absent, an LLM identifies latent error indicators from the
first-pass sample. This fallback is the only LLM call in the pipeline
that operates on raw log content; it is scoped to the bounded seed
sample ($10^4$ lines) and used solely for error-indicator
identification, not for RCA reasoning.

\subsection{Stages 2--3: Knowledge Retrieval and Schema Memory}
Before telemetry processing, the agent queries a Vector Knowledge
Base (KB) to retrieve \emph{Forensic Rules} associated with the
target log space---SRE-curated ground truth for success/failure
classification. These rules take absolute precedence over downstream
statistical analysis. For example: \texttt{flag='T'} implies success;
\texttt{return\_code='990000'} maps to ``Database Timeout''.

A few-shot Schema Inference module then maps heterogeneous raw columns
to a standardised Forensic Schema, identifying which column represents
the error code, log level, or success flag. Inferred schemas are
cached in a lightweight SQLite store (Schema Memory). Subsequent
analyses of the same log space retrieve the schema deterministically,
skipping the LLM call. In our deployment, this cache-aside pattern
reduces schema-inference cost to zero for all repeat analyses on
known services. Onboarding a new microservice requires only a
one-time KB rule registration from the responsible SRE team.

\subsection{Stage 4: Two-Layer Entropy-Guided Compression}
High-cardinality fields (e.g.\ UUIDs, request IDs) consume context
budget while contributing minimal diagnostic signal. We address this with
a two-layer entropy architecture.

\paragraph{Layer 1 (column-level).}
Shannon entropy
$H(C) = -\sum_v P(v) \log_2 P(v)$ is computed over the value
distribution of each metadata column $C$, and a compression decision
is applied:
\begin{equation}
\mathrm{compress}(C) =
\begin{cases}
\mathrm{ENUM}(C) & \text{if } |V_C| \le \tau_k \text{ or }
                   |V_C|/N < \tau_r,\\
\mathrm{ID}(C)   & \text{otherwise,}
\end{cases}
\label{eq:layer1}
\end{equation}
with $\tau_k{=}50$, $\tau_r{=}0.05$. Low-entropy fields (e.g.\
\texttt{return\_code} with $H{=}1.56$ bits) are retained as
categorical enumerations; high-entropy fields (e.g.\ randomised trace
IDs with $H{=}9.96$ bits) are abstracted to a uniqueness score and a
single sample. Column-wise analysis is parallelised via a map-reduce
pattern, allowing wide schemas (398 columns) to be processed within
the SLA.

\paragraph{Layer 2 (intra-template parameter-level).}
After Drain3~\cite{he2017drain} clustering, each log template contains
positional parameter slots. For each slot $s_j$, the system classifies
observed values:
\begin{equation}
\phi(s_j) = \begin{cases}
\mathrm{CONST} & \text{if } |V_{s_j}| = 1 \\
\mathrm{NUM}   & \text{if } > 90\% \text{ numeric} \\
\mathrm{ENUM}  & \text{if } |V_{s_j}| < 15 \\
\mathrm{ID}    & \text{otherwise.}
\end{cases}
\label{eq:layer2}
\end{equation}
Together, both layers reduce a stream of 3\,M raw log lines to a
stable 7{,}000--20{,}000-character compressed
context---a 1{,}000--7{,}000$\times$ reduction---independent of raw log
volume (\S\ref{sec:eval}).

\subsection{Stage 5: Contrastive Skew Analysis}
To provide statistically grounded evidence, the engine performs a
contrastive comparison between logs classified as errors and those
classified as successes. Error classification follows a strict
deterministic hierarchy: (1)~KB domain rules, (2)~explicit log level
(ERROR/FATAL), (3)~binary flag fields from Schema Inference,
(4)~keyword heuristics (\emph{exception}, \emph{timeout},
\emph{refused}).

For each categorical value $v$ in column $C$, a Critical Hint is
raised when the value is disproportionately concentrated in error logs:
\begin{equation}
\mathrm{Hint}(v,C) = \mathbf{1}\!\left(
P(v\mid \mathrm{Err}) > 0.01 \;\wedge\;
\frac{P(v\mid\mathrm{Err})}{P(v\mid\mathrm{Succ})+\varepsilon} > 3.0
\right),
\label{eq:hint}
\end{equation}
where $\varepsilon$ is a smoothing constant. Critical Hints are
injected into the final prompt with their exact probabilities,
bypassing text summarisation entirely. In the trace in
Table~\ref{tab:trace}, this automatically flagged
\texttt{clientVersion='12.11.30.303'} as appearing in 26.5\% of error
logs and 0\% of success logs.

\subsection{Stage 6: Orchestration and Generative Synthesis}
The Orchestrator assembles a Forensic Case File in a strict priority
order that ensures the highest-signal content is always placed at the
beginning of the context window, addressing the ``Lost in the Middle''
degradation~\cite{liu2024lost}:
\begin{enumerate}
  \item KB ground-truth rules (success/failure definitions),
  \item Ranked Critical Hints with exact probabilities,
  \item Compressed log templates (top-$n$, capped at 30 entries),
  \item Error timeline (auto-resampled from 5\,min to 10\,s resolution).
\end{enumerate}
The Orchestrator enforces a hard 46{,}000-character API budget. When
multiple log spaces are analysed (a common multi-service triage
scenario), the budget is distributed proportionally; statistical
sections always take priority over raw templates, which are truncated
from the bottom if necessary.

The LLM is instructed to act as a Senior SRE whose sole task is to
correlate the statistical hints with the semantic content of the log
templates---not to generate new facts. This constrained generation
makes synthesis-time confabulation harder: producing a hypothesis
absent from the evidence requires actively contradicting the injected
statistics rather than merely inferring from vague text.

\section{Industrial Evaluation}
\label{sec:eval}

\subsection{Experimental Setup}
\paragraph{Deployment environment.}
\textsc{Log-Insight} is deployed at Huawei against a ClickHouse
telemetry store under a 60-second end-to-end SLA and a
46{,}000-character LLM API budget per request. The system uses a
centralised general-purpose enterprise LLM accessed as a
service---deliberately avoiding bespoke fine-tuned models to minimise
MLOps overhead in a multi-tenant environment.

\paragraph{Dataset.}
The evaluation dataset comprises 11 historical incident scenarios
spanning 11 distinct log spaces (isolated, service-specific log
streams), ranging from 162 to 3{,}540{,}000 rows and 26 to 398
columns. Log-space names are anonymised for confidentiality.
Ground-truth labels were established post-incident: after
investigating and resolving each issue, the responsible SRE team
annotated the telemetry snapshot with a natural-language summary of
the true root cause.

\paragraph{Evaluation protocol.}
\textsc{Log-Insight} frames RCA as a retrieval task, so we measure
accuracy with Mean Reciprocal Rank (MRR). For each scenario, the
system produces a structured report from which we extract an ordered
list of distinct root-cause hypotheses. A secondary evaluator LLM
determines the rank at which the ground-truth label first appears in
the hypothesis list by checking for semantic entailment and paraphrase
equivalence. To validate this automated scoring, a human expert
independently scored a 20\% sample; raw inter-annotator agreement on
the assigned rank exceeded 90\%. Each scenario is run 10 times to
account for LLM stochasticity and sampling variance, yielding 110
total runs. The per-space MRR is:
\begin{equation}
\mathrm{MRR} = \frac{1}{|S|}\sum_{i=1}^{|S|}\frac{1}{R}\sum_{r=1}^{R}
   \frac{1}{\mathrm{rank}_{i,r}},
\label{eq:mrr}
\end{equation}
where $|S|{=}11$ log spaces and $R{=}10$ runs per space. Not-found
runs contribute $1/\mathrm{rank}{=}0$.

\subsection{Results}
Table~\ref{tab:per-space} reports per-log-space results.
\textsc{Log-Insight} achieves a macro-average MRR of 0.790 in
under a minute of end-to-end latency (mean 27\,s across all spaces). The correct root cause appears as the
top-ranked hypothesis in the majority of runs for six log spaces
(Pass); four spaces show mixed behaviour where the correct answer is
consistently identified but not always first (Mix); one space shows a
rank-2$+$ majority (Partial). The ground-truth label was not found in
only 4 of 110 runs.

\begin{table*}[t]
\caption{Per-log-space MRR results (mean over 10 independent runs).
Pass = rank 1 in majority of runs; Mix = ranks 1--3 with no dominant
position; Partial = rank 2$+$ in majority. Score $= 1/\mathrm{rank}$
averaged over 10 runs; not-found runs score 0 and are excluded from
Avg.\ Rank ($\dagger$). Macro-average MRR $=$ 0.790.}
\label{tab:per-space}
\small
\begin{tabular}{lccccccc}
\toprule
\textbf{Log Space} & \textbf{Status} & \textbf{MRR ($\pm$\,SD)} &
\textbf{Avg.\ Rank} & \textbf{Not Found} & \textbf{Time (s)} &
\textbf{Rows} & \textbf{Cols} \\
\midrule
space\_01 & Pass    & 0.883 $\pm$ 0.249 & 1.30          & 0 & 22.4 & 1{,}325       & 98 \\
space\_02 & Pass    & 0.700 $\pm$ 0.422 & 1.25$^\dagger$ & 2 & 19.8 & 1{,}495       & 398 \\
space\_03 & Mix     & 0.800 $\pm$ 0.258 & 1.40          & 0 & 31.2 & 500           & 98 \\
space\_04 & Mix     & 0.758 $\pm$ 0.320 & 1.70          & 0 & 24.7 & 452           & 98 \\
space\_05 & Partial & 0.567 $\pm$ 0.335 & 1.89$^\dagger$ & 1 & 38.5 & 1{,}000       & 98 \\
space\_06 & Mix     & 0.758 $\pm$ 0.320 & 1.70          & 0 & 21.3 & 10{,}000      & 98 \\
space\_07 & Mix     & 0.708 $\pm$ 0.317 & 1.80          & 0 & 27.6 & 100{,}000     & 98 \\
space\_08 & Pass    & 0.950 $\pm$ 0.158 & 1.10          & 0 & 23.1 & 189           & 26 \\
space\_09 & Pass    & 0.733 $\pm$ 0.370 & 1.44$^\dagger$ & 1 & 35.8 & 162           & 26 \\
space\_10 & Pass    & 0.883 $\pm$ 0.249 & 1.30          & 0 & 26.4 & 189           & 26 \\
space\_11 & Pass    & 0.950 $\pm$ 0.158 & 1.10          & 0 & 31.0 & 3{,}540{,}000 & 29 \\
\midrule
Aggregate & ---     & 0.790 $\pm$ 0.287 & 1.45          & 4 & 27.1 & ---           & --- \\
\bottomrule
\end{tabular}
\\[2pt]
$^\dagger$~Avg.\ Rank excludes not-found runs (score $= 0$).
\end{table*}

\subsection{Baseline Comparison}
Table~\ref{tab:baselines} compares \textsc{Log-Insight} against two
controls: \textbf{Random Sampling (RS)}, which injects a random subset
of raw logs directly into the LLM, and \textbf{Drain + Template
Sampling (DTS)}, which injects Drain-computed templates without
statistical scoring (i.e.\ the same Stage~6 orchestrator stripped of
Stages~4--5). For fairness, the sampling budget for both baselines
is constrained to the same log segments processed by
\textsc{Log-Insight}'s intelligent sampling. Evaluation uses
ROUGE-L, METEOR, BERT-based Semantic Similarity, and MRR, following
the protocol of~\cite{li2025coca}.

\textsc{Log-Insight} consistently outperforms both controls across
all metrics. The performance gap widens with log volume: on space\_11
(3.54\,M rows), \textsc{Log-Insight} achieves MRR~$=$~0.95 vs.\ 0.40
for DTS and 0.25 for RS. On small log spaces (space\_01, 1{,}325
rows), all methods converge: at low volume, intelligent sampling
alone provides sufficient error representation regardless of the
downstream processing.

\begin{table}[t]
\caption{Comparison against baseline methods (10 runs each on three
representative log spaces). Values are percentage means
($\pm$\,SD).}
\label{tab:baselines}
\small
\begin{tabular}{lcccc}
\toprule
\textbf{Method / Space} & \textbf{ROUGE-L} & \textbf{METEOR} &
\textbf{Sem.\ Sim.} & \textbf{MRR} \\
\midrule
\multicolumn{5}{l}{\emph{Random Sampling}} \\
space\_01 & 16.97 $\pm$ 1.36 & 21.90 $\pm$ 4.19 & 87.17 $\pm$ 0.47 & 0.80 \\
space\_07 &  3.47 $\pm$ 1.05 &  8.77 $\pm$ 1.48 & 81.78 $\pm$ 0.44 & 0.30 \\
space\_11 &  3.62 $\pm$ 2.75 &  7.66 $\pm$ 2.97 & 79.08 $\pm$ 0.14 & 0.25 \\
Avg.      &  8.02            & 12.78            & 82.68            & 0.45 \\
\midrule
\multicolumn{5}{l}{\emph{Drain + Template Sampling}} \\
space\_01 & 16.37 $\pm$ 0.67 & 22.25 $\pm$ 4.85 & 86.48 $\pm$ 1.05 & 0.80 \\
space\_07 &  4.29 $\pm$ 0.85 &  7.90 $\pm$ 0.27 & 81.75 $\pm$ 0.17 & 0.50 \\
space\_11 &  5.26 $\pm$ 1.12 &  8.56 $\pm$ 0.08 & 79.75 $\pm$ 0.27 & 0.40 \\
Avg.      &  8.64            & 12.90            & 82.66            & 0.56 \\
\midrule
\multicolumn{5}{l}{\emph{\textsc{Log-Insight} (ours)}} \\
space\_01 & 28.83 $\pm$ 1.10 & 34.36 $\pm$ 0.29 & 87.98 $\pm$ 0.22 & 0.88 \\
space\_07 & 15.99 $\pm$ 0.47 & 27.35 $\pm$ 1.48 & 86.00 $\pm$ 0.28 & 0.70 \\
space\_11 &  9.06 $\pm$ 0.84 & 20.75 $\pm$ 1.67 & 81.93 $\pm$ 0.14 & 0.95 \\
Avg.      & 17.96            & 27.49            & 85.30            & 0.84 \\
\bottomrule
\end{tabular}
\end{table}

BLEU scores (computed but omitted from the table for space) are
uniformly low across all methods due to creative rephrasing inherent
in LLM generation; BLEU penalises lexical divergence regardless of
semantic accuracy, making it an unreliable indicator for open-ended
RCA generation. Semantic similarity scores are uniformly high
(79--88\%) across all methods, indicating that all methods produce
responses in the correct semantic neighbourhood; the key
discriminator is whether the specific root cause is identified and
ranked first, which MRR captures directly.

\subsection{Context Compression}
\begin{figure}[t]
\centering
\begin{tikzpicture}
\begin{axis}[
  width=0.95\columnwidth, height=4.6cm,
  xlabel={Number of raw log lines},
  ylabel={Characters},
  xmin=0, xmax=3000000, ymin=0, ymax=50000,
  xtick={0,1000000,2000000,3000000},
  xticklabels={0,1M,2M,3M},
  ytick={0,10000,20000,30000,46000},
  yticklabels={0,10k,20k,30k,46k},
  legend style={
    at={(0.97,0.62)}, anchor=east,
    font=\scriptsize, draw=gray!50,
    fill=white, fill opacity=0.9,
    text opacity=1, legend cell align=left,
  },
  tick label style={font=\scriptsize},
  label style={font=\scriptsize},
  grid=major, grid style={dashed,gray!30},
]
\addplot[red, thick, dashed, domain=0:80] {600*x};
\addplot[red, thick, dotted] coordinates {(0,46000) (3000000,46000)};
\addplot[blue, thick] coordinates {(0,7000) (1000000,12000) (3000000,18000)};
\legend{Raw injection (hits limit), 46k API ceiling, \textsc{Log-Insight} (compressed)}
\end{axis}
\end{tikzpicture}
\caption{Character consumption vs.\ raw log count. Raw injection
exhausts the 46k API limit at $\approx$50--80 lines.
\textsc{Log-Insight}'s compressed context stays stable at
7k--20k characters across streams up to 3\,M logs
(1{,}000--7{,}000$\times$ reduction).}
\label{fig:compression}
\end{figure}
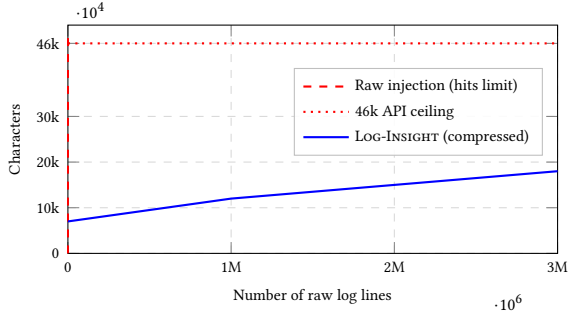

The API character limit is the primary bottleneck for direct LLM-based log analysis. Figure~\ref{fig:compression} illustrates this:
raw-log injection exhausts the 46{,}000-character ceiling at
approximately 50--80 log lines, rendering direct injection infeasible
at any production scale. \textsc{Log-Insight}'s two-layer entropy
compression keeps the compressed context stable at
7{,}000--20{,}000 characters across streams of up to 3\,M lines---a
1{,}000--7{,}000$\times$ reduction---while preserving critical
statistical signals. For space\_11 (3.54\,M rows), the uncompressed
baseline would require approximately 3.7 billion characters; the
compressed context is approximately 18{,}000 characters.

\subsection{Failure-Mode Analysis and Mitigations}
The evaluation reveals two failure modes.

\paragraph{Context omission.}
In log spaces with unusually wide schemas (space\_02, 398 columns),
the character budget is consumed by metadata enumeration before the
contrastive skew report can be fully injected. This deprives the
LLM of its statistical anchor and accounts for 2 of the 4 not-found
runs (both occurring in space\_02). The mitigation under
investigation is adaptive budget allocation: dynamically prioritising
the skew section over metadata enumeration when column count exceeds
a threshold.

\paragraph{Synthesis failure.}
The remaining 2 not-found runs (in space\_05 and space\_09) and the
rank-3+ outcomes in spaces 04, 06, and 07 reflect a different
pattern: the deterministic preprocessing stages correctly isolate
and inject the true root-cause signal, yet the LLM assigns it lower
narrative priority, typically when the anomaly is a domain-specific
error code that conflicts with a more ``interesting'' causal chain
in the model's parametric memory. Two mitigations are being tested:
(1)~structured-output enforcement, requiring the LLM to first return
a JSON evidence-to-claim mapping before generating free text; and
(2)~a critic-agent pass, where a secondary LLM validates that the
final narrative ordering strictly reflects the injected skew
statistics.

\subsection{Human-in-the-Loop and Adoption}
Deploying LLMs for live incident response requires SRE trust. During the initial deployment, engineers were
understandably hesitant to act on generative summaries under
high-stakes outage conditions. \textsc{Log-Insight} addresses this
through transparent design: every RCA report includes a
\emph{Forensic Evidence section} listing the exact log templates and
skew percentages used to form each hypothesis. Engineers can
independently verify this evidence in ClickHouse before acting.

The structure of the Forensic Evidence section is explicit by
design. For each Critical Hint, the report states: the field name and
value, the exact percentage of error logs containing that value, the
exact percentage of success logs containing it, and the skew ratio.
For example: ``\texttt{hostAddr='10.140.111.13'}: present in 15.3\% of
error logs vs.\ 0.0\% of success logs (skew ratio: $\infty$).'' This
is not a summary or interpretation---it is the raw output of
Stage~5, reproduced verbatim. SREs can query ClickHouse directly to
verify these numbers in under 30 seconds, creating a verifiable
audit trail between system conclusions and raw data.

Qualitative feedback from the operations team during the deployment
period consistently identified statistical transparency as the
primary factor enabling adoption. Specifically, engineers reported that the
ability to verify individual hints before acting on the full report
is the key enabling factor---even when they trust the aggregate
conclusion, they use the Forensic Evidence section to build
intuition about the incident and identify follow-up investigation
steps. This aligns with recent empirical studies on trace
analysis adoption~\cite{shahedi2025excellence}, which find that
transparency and expert-knowledge embedding distinguish
tools that are technically excellent from those that are
actually adopted in practice.

\section{Discussion}

\subsection{Generalisability of the Neuro-Symbolic Design}
The core architectural principle---separating high-volume symbolic
preprocessing from constrained neural synthesis---is not specific to
log analysis. It applies to any domain where an LLM must reason
over high-volume, low-signal operational data that exceeds its
context window. The same pattern is applicable to distributed
tracing data (spatiotemporal chain analysis before LLM synthesis),
configuration-diff analysis (entropy filtering of unchanged fields
before reasoning about the changed ones), and security audit log
analysis (skew detection between anomalous and normal sessions
before LLM synthesis).

The design also generalises across compression technologies. Drain3
provides template clustering for log message bodies; Shannon entropy
provides column-level filtering. Neither is essential to the
architecture: any clustering algorithm producing countable, typed
events can substitute for Drain3, and any information-theoretic
ranking criterion can substitute for entropy. The key invariant is
the separation of concerns: symbolic stages rank by diagnostic
relevance, neural stages synthesise from ranked evidence.

\subsection{Lessons for the SE Community}

\textbf{Statistical grounding limits LLM hallucination
structurally.} Instructing the LLM to ``only use evidence from the
provided context'' via prompt engineering is insufficient in
practice: models still draw on parametric memory when the evidence
is ambiguous. As described in \S\ref{sec:system}, injecting
probability-annotated evidence makes confabulation harder because
it requires actively contradicting a stated statistic.

\textbf{Latency is dominated by symbolic, not neural, stages.}
Counter-intuitively, the dominant cost driver in
\textsc{Log-Insight} is not the LLM API call but the Drain3 parsing
step. Optimising the LLM (e.g.\ switching to a faster model) has
less impact on end-to-end latency than optimising the symbolic
preprocessing. For future work targeting sub-10-second latency,
the symbolic pipeline is the bottleneck.

\subsection{Limitations and Open Directions}
\label{sec:limitations}

This work is an initial deployment study on 11 incidents from a
single operational environment. We identify the following
limitations.

\paragraph{KB dependency.}
The current system requires a one-time SRE rule registration per
log space. While this is a low barrier in practice (a single
annotation per microservice), it limits zero-shot generalisation
to entirely new services. Automated KB bootstrapping from
historical incident post-mortems could remove this requirement.

\paragraph{Threshold sensitivity.}
The skew-analysis thresholds ($P(v|\mathrm{Err}) > 0.01$, ratio
$> 3.0$) and entropy compression thresholds ($\tau_k{=}50$,
$\tau_r{=}0.05$) were calibrated empirically for our deployment
environment. A systematic sensitivity analysis across diverse
log schemas remains open.

\paragraph{Baseline completeness.}
The current evaluation compares against random sampling and Drain
template injection. A naive RAG baseline---dense retrieval over
raw log chunks using embedding similarity---would directly isolate
the contribution of symbolic preprocessing over semantic retrieval.
We leave this comparison to future work.

\paragraph{Infrastructure and model dependency.}
All experiments ran on CPU-only infrastructure using a
general-purpose enterprise LLM. Fine-tuned Small Language Models
specialised for log semantics could reduce both latency and cost.
A locally-deployed SLM would also address data-residency concerns
that prevent some teams from using cloud LLM APIs for live
incident analysis.

\section{Threats to Validity}

\paragraph{Internal validity.}
The evaluation uses an LLM-as-a-judge to score generated hypotheses
against SRE ground truth. While raw inter-annotator agreement with
human experts exceeded 90\% on a 20\% sample, the judge LLM could
introduce systematic biases---for example, preferring hypotheses
that share vocabulary with the ground truth even when the
underlying claim is incorrect. Reporting Cohen's
$\kappa$ against a larger human-annotated sample would
strengthen confidence in the evaluation.

The 10-run repetition per log space accounts for LLM generation
stochasticity but does not fully account for sampling variance from
Stage~1. In log spaces with extremely rare error templates
(below 0.1\% occurrence rate), sampling variance could affect which
templates are included in downstream analysis.

\paragraph{External validity.}
The evaluation covers 11 incident scenarios from a single
organisation's microservice environment, all using the same
ClickHouse telemetry backend. The results may not generalise to
organisations using different log architectures (e.g.\ ElasticSearch
stacks, unstructured flat-file logs) or incident types that do not
manifest as distributional skew in categorical fields. Incidents
caused by performance degradation (high-latency rather than error
spikes) may require different statistical detection strategies.

\paragraph{Construct validity.}
MRR measures the rank of the ground-truth root cause in the
generated hypothesis list, which approximates diagnostic utility
for an SRE. However, it does not capture other aspects of report
quality such as the accuracy of supporting evidence or the
actionability of proposed next investigation steps.

\section{Conclusion}

\textsc{Log-Insight} demonstrates that automating the structured
investigation workflow of a skilled SRE---through neuro-symbolic
log compression, statistical anomaly ranking, and constrained LLM
synthesis---is a viable path to scalable incident diagnosis in
production microservice environments. Across 110 runs on 11 real
production incidents, it achieves MRR~$=$~0.790 in under a minute of
latency and up to 7{,}000$\times$ context compression, handling
log spaces of 3.54\,M rows within a fixed API budget.

The deployment experience yields three lessons with broader
applicability. First, statistical transparency is a first-class
engineering requirement. Second, the Lost-in-the-Middle problem
can be structurally avoided by ensuring that the highest-signal
content is always placed at the beginning of the LLM context.
Third, symbolic preprocessing changes the nature of the LLM task:
the system is not asking the LLM to read raw data and infer
patterns, but to correlate two already-ranked pieces of evidence
(statistical skew and log-template semantics)---a substantially
easier cognitive task that explains why a general-purpose LLM
with no log-specific fine-tuning achieves strong RCA performance.

The principled separation of symbolic and neural computation in
\textsc{Log-Insight} suggests a general design pattern for SE
automation tools that operate over large, noisy operational
datasets: invest in rigorous symbolic preprocessing to reduce the
problem to a well-posed reasoning task, then use LLMs for their
reasoning and generation strengths. The success of this approach
in the incident-diagnosis domain motivates exploring it in
adjacent problems: automated test-failure triage, deployment
configuration analysis, and performance-regression
investigation.

\section*{Data Availability Statement}

The evaluation telemetry analysed in this paper consists of
production logs from Huawei's internal microservice infrastructure,
including customer requests, service identifiers, and internal
component behaviour. This data is subject to commercial
confidentiality obligations and customer-data protection
requirements that prevent its public release in any form, redacted
or otherwise. The 11 log spaces, their schemas, ground-truth
annotations, and raw telemetry therefore cannot be archived
publicly.

The \textsc{Log-Insight} pipeline implementation is currently
deployed and under active evaluation in this production
environment. Releasing the implementation publicly at this stage
would risk exposing internal service-schema assumptions and
SRE-curated Knowledge Base rules that are tightly coupled to
specific Huawei systems and customer-facing services. We therefore
do not release the production codebase as part of this submission.

We plan to revisit a public artifact release once the production
deployment stabilises and legal review of derivative
artifacts is complete.

\begin{acks}
The authors thank the HQ SRE Operations Team at Huawei for their
invaluable operational feedback and for providing the production
telemetry required to validate this framework. This work was
supported by Huawei, which sponsored the research and enabled the
deployment of \textsc{Log-Insight} in a large-scale production
environment.
\end{acks}

\bibliographystyle{ACM-Reference-Format}

\end{document}